\documentclass[prb,aps,nofootinbib,amssymb,twocolumn,superscriptaddress,10pt]{revtex4-2}
\usepackage[T1]{fontenc}
\usepackage{multirow}
\usepackage{bbold}
\usepackage{color}
\usepackage{xcolor}
\usepackage{mathrsfs}
\usepackage{hyperref}
\usepackage{bm}
\usepackage{stackrel}

\usepackage{amsfonts, relsize, color}
\usepackage{graphics}
\usepackage{graphicx}
\usepackage{amssymb}
\usepackage{amsbsy}
\usepackage{amsmath}
\usepackage{array}
\usepackage{fontenc}
\usepackage{textcomp}
\usepackage{rotating}
\usepackage{comment}
\usepackage{physics}
\usepackage{braket}
\usepackage[normalem]{ulem}
\usepackage{cleveref}

\usepackage{float,latexsym}
\usepackage{multirow}
\usepackage{tikz}
\usepackage{color}
\usepackage{xr}
\usepackage{siunitx}
\usepackage[caption=false]{subfig}

\crefname{appendix}{Appendix}{Appendices}
\crefname{equation}{Eq.}{Eqs.}
\crefname{figure}{Fig.}{Figs.}
\crefname{table}{Table}{Tables}
\crefname{section}{Section}{Sections}
\crefname{enumi}{Case}{Cases}
\creflabelformat{appendix}{[#2#1#3]}
\crefrangelabelformat{figure}{#3#1#4--(#5\crefstripprefix{#1}{#2}#6}
\crefmultiformat{figure}{Figs.~#2#1\xdef\mycreffirstarg{#1}#3}%
 { and~#2#1#3}{, #2#1#3}{ and~#2#1#3}
\newcommand{\ie}{{\it i.e.}}
\newcommand{\eg}{{\it e.g.}}

\begin{document}
\title{Thermal spin dynamics of Kitaev magnets  ---  scattering continua and \\magnetic field induced phases within a stochastic semiclassical approach}

\author{Oliver Franke}
\thanks{These authors contributed equally.}%
\affiliation{Department of Physics TQM, Technische Universit{\"a}t M{\"u}nchen, James-Franck-Stra{\ss}e 1, D-85748 Garching, Germany}%
\affiliation{Dahlem Center for Complex Quantum Systems and Physics Department, Freie Universit{\"a}t Berlin, Arnimallee 14, 14195 Berlin, Germany}%

\author{Dumitru C\u{a}lug\u{a}ru}%
\thanks{These authors contributed equally.}%
\affiliation{Department of Physics, Princeton University, NJ 08544, USA}%
\affiliation{Cavendish Laboratory, University of Cambridge, Cambridge, CB3 0HE, United Kingdom}%

\author{Andreas Nunnenkamp}%
\affiliation{Faculty of Physics, University of Vienna, Boltzmanngasse 5, 1090 Vienna, Austria}
\affiliation{Cavendish Laboratory, University of Cambridge, Cambridge, CB3 0HE, United Kingdom}

\author{Johannes Knolle}%
\affiliation{Department of Physics TQM, Technische Universit{\"a}t M{\"u}nchen, James-Franck-Stra{\ss}e 1, D-85748 Garching, Germany}%
\affiliation{Munich Center for Quantum Science and Technology (MCQST), 80799 Munich, Germany}%
\affiliation{Blackett Laboratory, Imperial College London, London SW7 2AZ, United Kingdom}

\date{\today}

\begin{abstract}
The honeycomb magnet $\alpha-$RuCl$_3$ is a prime candidate material for realizing the Kitaev quantum spin liquid (QSL), but it shows long-range magnetic order at low temperature. Nevertheless, its broad inelastic neutron scattering (INS) response at finite frequency has been interpreted as that of a 'proximate QSL'. A moderate in-plane magnetic field  indeed melts the residual zigzag order, giving rise to peculiar intermediate field phases before the high-field polarized state. In INS measurements the low-frequency spin waves disappear, leading to a broad scattering continuum in the field-induced intermediate regime, whose nature is currently under debate. Here, we study the magnetic field dependent spin dynamics of the  $K-\Gamma-\Gamma'-$model within a stochastic semiclassical treatment, which incorporates the effect of finite-temperature fluctuations. At temperatures relevant for INS experiments, we show how the excitations of the zigzag phase broaden and that the different intermediate phases all show a similar continuum response. We discuss the implications of our results for experiments and highlight the importance of distinguishing finite temperature fluctuations from genuine quantum fractionalization signatures in frustrated magnets.
\end{abstract}

\maketitle

\section{Introduction}\label{sec:introduction}
The dynamical spin structure factor, as measured in inelastic neutron scattering (INS), is an ideal tool for gaining a comprehensive understanding of quantum magnets. However, in sought-after quantum spin liquids (QSL)~\cite{savary2016quantum,knolle2019field,broholm2020quantum}, it only yields a broad continuum response~\cite{han2012fractionalized}, as local spin flip excitations decay into multiple fractionalized excitations. This has complicated the unambiguous identification of a genuine QSL in a material. 

In recent years, a number of Mott insulating systems with strong spin orbit coupling~\cite{jackeli2009mott} have been put forward as candidates for realising the Kitaev honeycomb QSL~\cite{kitaev2006anyons} (see Refs.~\cite{hermanns2018physics,winter2017models,takagi2019concept,motome2020hunting,trebst2022kitaev} for reviews on the subject). The layered honeycomb magnet $\alpha-$RuCl$_3$~\cite{plumb2014alpha} appears to be one of the most promising, as its INS~\cite{banerjee2016proximate,banerjee2017neutron,do2017majorana} and Raman spectroscopic response~\cite{sandilands2015scattering,nasu2016fermionic,wang2020range} display broad scattering continua at elevated frequency similar to the predictions for the ideal Kitaev model~\cite{KNO14,KNO15,yoshitake2016fractional,yoshitake2017temperature,knolle2018dynamicsb,knolle2014raman}. Despite the presence of long-range zigzag magnetic order at low temperature~\cite{sears2015magnetic,johnson2015monoclinic}, these have been interpreted as signatures of fractionalized excitations of a 'proximate QSL' nearby in the phase diagram~\cite{hermanns2018physics}. Indeed, via the application of a moderate in-plane magnetic field of about \SI{8}{\tesla}~\cite{kubota2015successive,sears2015magnetic,sears2017phase,wolter2017field}, the zigzag order is suppressed, giving rise to unusual intermediate field phases with a whole range of atypical properties~\cite{janssen2019heisenberg}. For example, a thermal Hall response reminiscent of the one predicted for the non-Abelian Kitaev QSL~\cite{kitaev2006anyons} has been reported in Refs.~\cite{kasahara2018majorana,yokoi2021half}, which is currently under debate~\cite{bruin2022robustness,yamashita2020sample,lefranccois2022evidence,czajka2022planar}. 

INS measurements in a field have shown how the low frequency spin wave excitations of $\alpha-$RuCl$_3$ melt as the zigzag order disappears for increasing magnetic fields giving rise to a broad scattering continuum centred around the $\Gamma$-point of the Brillouin zone~\cite{BAN18}. Whether the broad scattering response is best understood as a signature of (weakly confined) fractional spin excitations or arises due to nonlinearities beyond harmonic magnon dynamics, for example from magnon-magnon interactions, is again currently under debate~\cite{winter2018probing,winter2017breakdown,wang2017magnetic,little2017antiferromagnetic,wu2018field,sahasrabudhe2020high,shi2018field}. 

A further complication arises because the microscopic Hamiltonian describing the low-energy magnetic degrees of freedom of $\alpha-$RuCl$_3$ is, as of yet, not known. A consensus has been reached that a strong bond-dependent Kitaev exchange is present, but the value of the perturbing Heisenberg and spin-off-diagonal $\Gamma$ and $\Gamma'$ interactions (and further neighbor interactions) remains under discussion~\cite{kim2016crystal,eichstaedt2019deriving,maksimov2020rethinking,li2021identification}. In that context, the observation of a sub-leading yet sizeable out-of-plane modulation of excitations in INS~\cite{balz2021field} points to the importance of interlayer couplings. Nevertheless, an extended $K-\Gamma-\Gamma'-$model can capture a number of qualitative features of $\alpha-$RuCl$_3$ including magnetic-field induced intermediate phases~\cite{gordon2019theory,hickey2019emergence}. For example, Ref.~\cite{CHE20a} showed that, within a classical Monte Carlo sampling scheme, a whole zoo of intermediate phases appears between the low-field zigzag phase and the high-field polarized state~\cite{liu2021revealing,rao2021machine,chern2021classical}. Classically, these are characterized as non-collinear/coplanar states with large magnetic unit cells and their weak long-range order is expected to be unstable upon the inclusion of quantum and/or thermal fluctuations~\cite{lee2020magnetic}. 

Here, we study the temperature dependent spin excitations of the $K-\Gamma-\Gamma'-$model as a function of magnetic field. We employ a stochastic semiclassical method~\cite{eriksson2017atomistic} which we show can reproduce the intermediate field phases found via classical Monte Carlo sampling~\cite{CHE20a}. In addition, our method incorporates the effect of thermal fluctuations on the dynamical response~\cite{skubic2008method}, relevant for interpreting INS experiments. So far, these have been carried out for temperatures considerably lower, yet as we argue not low enough, than the bare exchange scales (Ref.~\cite{BAN18} reports e.g.~results down to approx.~$\SI{2}{\kelvin}$ with an estimate of the Kitaev exchange of approx.~$\SI{100}{\kelvin}$). Moreover, most theoretical modeling has been restricted to zero temperature quantum calculations except for a few recent exceptions~\cite{yoshitake2017temperature,nasu2016fermionic,rousochatzakis2019quantum}. 

Our choice of method is motivated by the remarkable finding of Ref.~\cite{SAM17} that the semiclassical Landau-Lifshitz dynamics (starting from initial states which are sampled via a low-temperature classical Monte Carlo approach) can capture the salient features of the pure Kitaev model. Concretely, the broad frequency continua and weak momentum modulation of the classical Kitaev spin liquid is remarkably similar to the one of the exact QSL at zero temperature. Only the low frequency response differs as it is governed by quantum selection rules associated with fractionalized flux excitations~\cite{KNO14,knolle2016dynamics}. Moreover, semiclassical dynamics of thermally disordered frustrated magnets have recently been shown to capture the INS response of other QSL candidates like NaCaNi$_2$F$_7$~\cite{zhang2019dynamical}, MgCr$_2$O$_4$~\cite{bai2019magnetic}, Ce$_2$Zr$_2$O$_7$~\cite{smith2022case, BHA22} or the $\Gamma$-model~\cite{samarakoon2018classical}. Thus, our work similarly addresses more general questions beyond the concrete example of $\alpha-$RuCl$_3$, namely understanding the broad INS scattering continua of frustrated magnets, and diagnosing genuine quantum fractionalization signatures. 

\section{Model, method and phase diagram}\label{sec:model}
\begin{figure}[t]
	\centering
	\includegraphics[trim=5 10 25 30, clip, width=1.0\columnwidth]{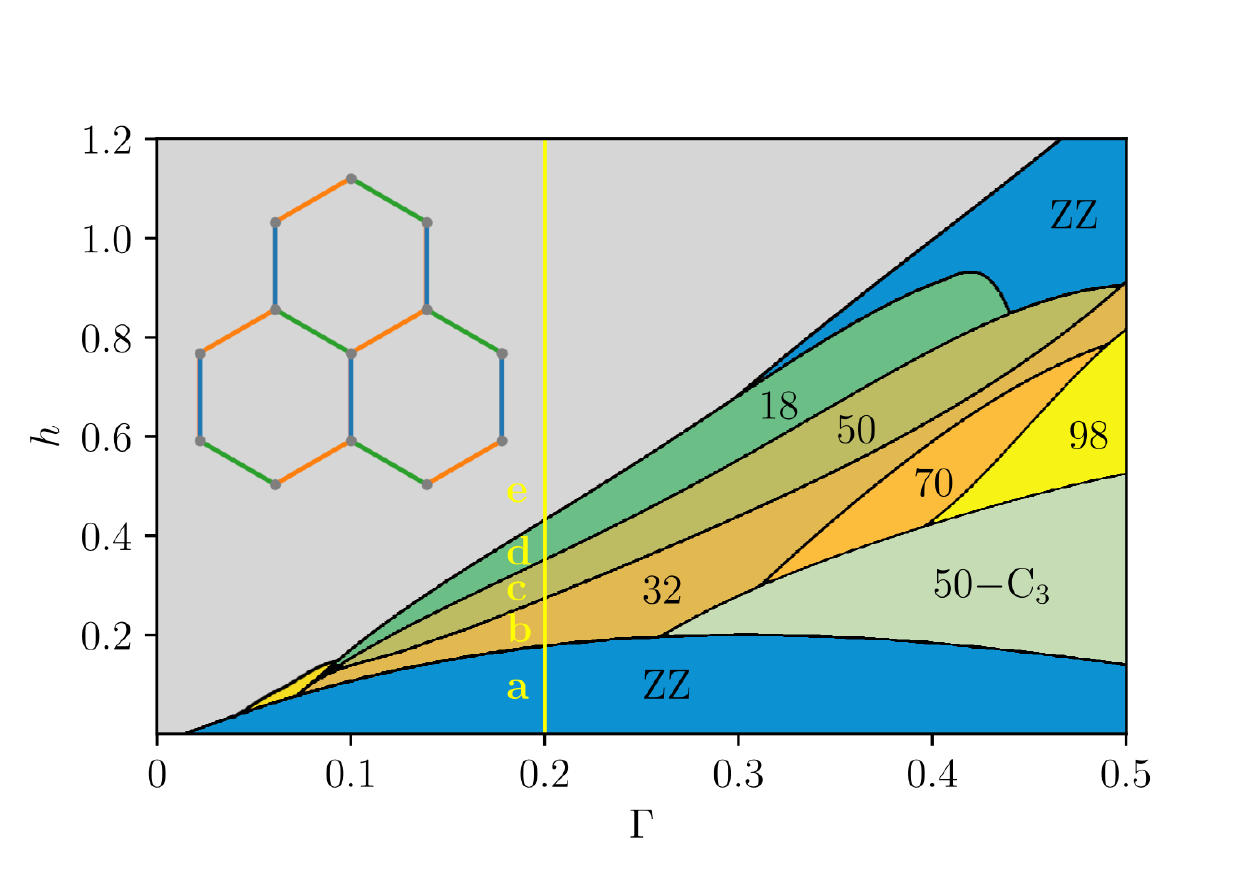}%
	\subfloat{\label{fig:phase_diagram_1:a}}%
    \subfloat{\label{fig:phase_diagram_1:b}}%
    \subfloat{\label{fig:phase_diagram_1:c}}%
    \subfloat{\label{fig:phase_diagram_1:d}}% 
    \subfloat{\label{fig:phase_diagram_1:e}}%
	\caption{Phase diagram of the field induced phases of the $K-\Gamma-\Gamma'-$model. The magnetic orders dependent on the $\Gamma$ interaction and the $[111]$-oriented magnetic field $h$. We identify the orders along the $\Gamma = 0.2$ line by their static structure factors (see \cref{fig:staticAll}) and present their dynamic structure factors at different temperatures in \cref{fig:dynamicAll}. The inset shows the honeycomb lattice with bond-dependent exchange interactions, \textit{i.e.}, each type of bond $\lambda \in \left\lbrace \mathrm{x}, \mathrm{y}, \mathrm{z} \right\rbrace$ is represented by a different color. Figure adapted from Ref. \cite{CHE20a}.}
	\label{fig:phase_diagram_1}
\end{figure}

We describe the honeycomb magnet $\alpha-$RuCl$_3$ within a \mbox{$K-\Gamma-\Gamma'-$model} and focus on the Hamiltonian studied in Ref.~\cite{rau2014generic,CHE20a}
\begin{equation}
    \label{eqn:Hamiltonian}
\begin{split}
    \mathcal{H} &= \sum_{\lambda = x,y,z} \sum_{\braket{ij} \in \lambda} \big[ K S_i^\lambda S_j^\lambda + \Gamma \left( S_i^\mu S_j^\nu + S_i^\nu S_j^\mu \right) \\
    &+ \Gamma^\prime \left( S_i^\mu S_j^\lambda + S_i^\lambda S_j^\mu + S_i^\nu S_j^\lambda + S_i^\lambda S_j^\nu \right) \big] - \mathbf{h} \cdot \sum_i \mathbf{S}_i,
\end{split}
\end{equation}
which includes the bond-dependent ferromagnetic Kitaev interaction, $K<0$, and the off-diagonal $\Gamma$ and $\Gamma^\prime$ interactions. In \cref{eqn:Hamiltonian}, $\braket{ij} \in \lambda$ denotes the nearest-neighbor pair formed by the spins $\mathbf{S}_i$ and $\mathbf{S}_j$ with bond orientation $\lambda \in \left\lbrace \mathrm{x}, \mathrm{y}, \mathrm{z} \right\rbrace$, as shown in the inset of \cref{fig:phase_diagram_1}. The off-diagonal interactions are given by cyclic permutations of the spin components $(\lambda, \mu, \nu)$. In the following, we will set $K=-1$ and use dimensionless parameters in units of $|K|$ and fundamental physical constants, \textit{e.g.} frequency $\omega$ is implicitly expressed in units of $|K|/ \hbar$ and the temperature $T$ in units of $|K|/k_B$.

%%%%%%%%%%%%%%%%%%%%%%%%%%%%%%%%%%%%%%%%%%%%%%%%%%%%%%%%%%%%%%
%%%%%%%%%%%%%%%%%%%%%%%%%%%%%%%%%%%%%%%%%%%%%%%%%%%%%%%%%%%%%%
%%%%%%%%%%%%%%%%%%%%%%%%%%%%%%%%%%%%%%%%%%%%%%%%%%%%%%%%%%%%%%
%%%%%%%%%%%%%%%%%%%%%%%%%%%%%%%%%%%%%%%%%%%%%%%%%%%%%%%%%%%%%%

The classical Kitaev spin liquid quickly reaches a polarized state in a $[111]$-oriented magnetic field $\mathbf{h} \neq 0$, but a finite $\Gamma$ interaction leads to a multitude of intermediate magnetic field induced ordered phases, which posses large unit cells and persist to greater values of $h=\abs{\mathbf{h}}$ for increasing $\Gamma$. Even more fragile magnetic orders are realised when adding a small $\Gamma^\prime = -0.02$ interaction, which also stabilises the zigzag phase around $h \approx 0$. The phase diagram as a function of $\Gamma$ and $[111]$-oriented magnetic field is depicted in \cref{fig:phase_diagram_1}, with the different orders named according to the number of spins in their respective unit cells (adapted from Ref. \cite{CHE20a}).

%%%%%%%%%%%%%%%%%%%%%%%%%%%%%%%%%%%%%%%%%%%%%%%%%%%%%%%%%%%%%%
%%%%%%%%%%%%%%%%%%%%%%%%%%%%%%%%%%%%%%%%%%%%%%%%%%%%%%%%%%%%%%
\begin{figure*}[t]
	\centering
	\includegraphics[width=2.0\columnwidth]{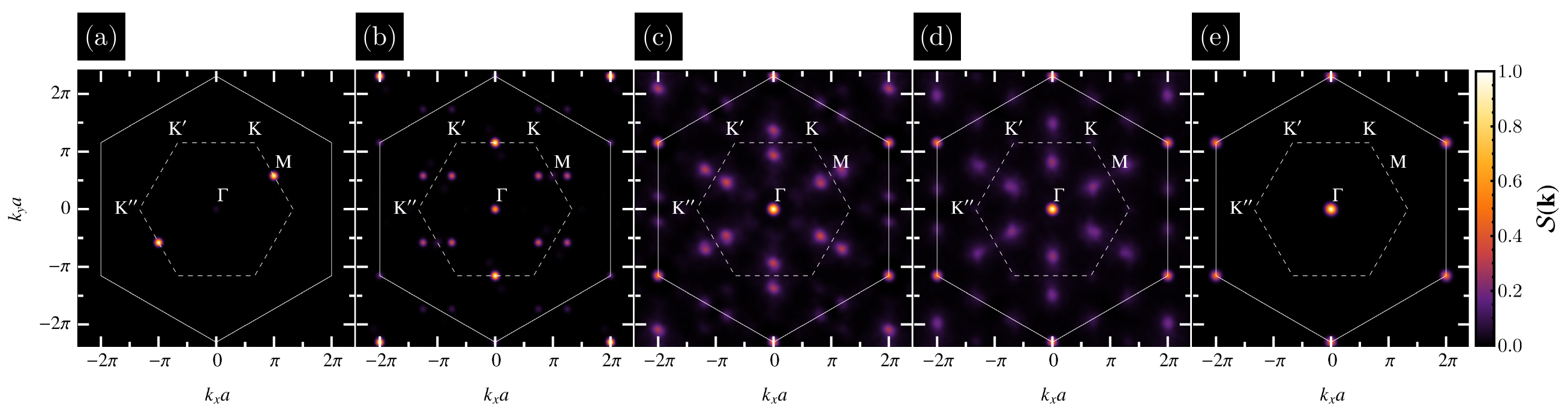}
	\caption{Static structure factors along the $\Gamma = 0.2$ line of the phase diagram in \cref{fig:phase_diagram_1} at low temperature $T=0.0001$. The sharp peaks are broadened by a Gaussian filter ($\sigma = 0.12$) for visibility and the first Brillouin zone is indicated by a dashed line. For concreteness, we show results for the points $|\mathbf{h}| \in \{0.1,0.225,0.35,0.386,0.6\}$ denoted by (a) to (e) in \cref{fig:phase_diagram_1} with $\Gamma^\prime = -0.02$. The structure factors correspond to single spin configurations, of which the ZZ and 32-site order are not $\mathcal{C}_3$-symmetric. Note that we normalized the structure factors individually to a maximum of one and scaled the intensity of (c) to (e) as $\sqrt{S(\mathbf{k})}$ to enhance the visibility of the peaks.}
	\label{fig:staticAll}
\end{figure*}
%%%%%%%%%%%%%%%%%%%%%%%%%%%%%%%%%%%%%%%%%%%%%%%%%%%%%%%%%%%%%%
%%%%%%%%%%%%%%%%%%%%%%%%%%%%%%%%%%%%%%%%%%%%%%%%%%%%%%%%%%%%%%
We study the spin dynamics based on the atomistic Landau-Lifshitz-Gilbert (LLG) equation
\begin{equation}
    \frac{\partial \mathbf{S}_i}{\partial t} = \frac{- 1}{(1 + \alpha^2)} \left[ \mathbf{S}_i \times \mathbf{H}_i(t) + \alpha \mathbf{S}_i \times (\mathbf{S}_i \times \mathbf{H}_i(t)) \right],
    \label{eq:LLG}
\end{equation}
which describes the damped precession of the classical spins around a local effective (exchange) field
\begin{equation}
    \mathbf{H}_i(t) = -\frac{\partial \mathcal{H}_i (\mathbf{S}_i(t))}{\partial \mathbf{S}_i} + \mathbf{b}_i(t).
    \label{eq:Heff}
\end{equation}
Here, the spins are represented by their normalized magnetic moments $\mathbf{S}_i$ at site $i$ (with $|\mathbf{S}_i|=1$). An effective damping of the dynamics from coupling to lattice and other degrees of freedom is included via the dimensionless parameter $\alpha$~\cite{skubic2008method}. For concreteness, we  fix it to a small nonzero value $\alpha = 0.0075$ to incorporate both fluctuations and dissipation whilst allowing for the propagation of long-lived spin waves~\cite{eriksson2017atomistic}.

We include the effects of finite temperature via a stochastic magnetic field $\mathbf{b}_i(t)$. This thermal noise, which describes the interaction of the system with a thermostat (\eg{} of a phonon subsystem) obeys
\begin{align}
    \braket{\mathbf{b}_i(t)}&= 0 \\
    \braket{\mathrm{b}_i^\nu(t) \mathrm{b}_j^\kappa(t')} &= 2 \alpha T \delta_{ij} \delta_{\nu \kappa} \delta(t-t')
    \label{eq:fdt}
\end{align}
for the three components of the spins $\nu, \kappa = \mathrm{x}, \mathrm{y}, \mathrm{z}$. The definition of the stochastic field ensures thermodynamic consistency as it reproduces a stationary Boltzmann probability distribution of the magnetic moments in statistical equilibrium~\cite{skubic2008method}. The approximation of uncorrelated ``white noise'' is justified when the autocorrelation time of the stochastic field is much shorter than the response of the system. Although this assumption breaks down at very low temperatures~\cite{Barker2019}, it is an efficient way of including the general qualitative effects of thermal fluctuations on the dynamical magnetic response~\cite{skubic2008method}.

We solve the system of nonlinear coupled stochastic differential equations, \cref{eq:LLG}, with an adaptive Runge-Kutta (RK) method of 4th order~\cite{DOR80} on Graphics Processing Units (GPUs) using the parallel computing platform CUDA. The parallel architecture allows us to significantly improve performance at two major steps, namely the calculation of the effective magnetic field $\mathbf{H}_i(t)$ using sparse matrix-vector multiplications and the local spin updates according to \cref{eq:LLG}, whereby each spin component is mapped to a GPU thread. This enables us to explore systems of up to $14 \, 112$ spins ($84 \times 84$ unit cells). More details about the adaptive RK method employed in this work are provided in \cref{app:sec:RK_method}.

We determine the classical ground state of the system using a simulated annealing prescription (see \cref{app:sec:sim_annealing}). In short, we start with a ferromagnetic initial state at high temperature (in which the spins are pointed along the $[111]$ direction) and then repeatedly cool and reheat the system until the final temperature is reached. The annealing process takes about $t = 10^7-10^8$ time units.

Our stochastic LLG method is in agreement with the phase diagram of Ref.~\cite{CHE20a}, which was obtained previously via a standard classical Monte Carlo sampling. We identify the competing intermediate orders along the \mbox{$\Gamma = 0.2$} line via the real space spin configurations and by their static structure factors shown in \cref{fig:staticAll}.

%%%%%%%%%%%%%%%%%%%%%%%%%%%%%%%%%%%%%%%%%%%%%%%%%%%%%%%%%%%%%%
%%%%%%%%%%%%%%%%%%%%%%%%%%%%%%%%%%%%%%%%%%%%%%%%%%%%%%%%%%%%%%
\begin{figure*}[t!]
	\centering
	\includegraphics[width=2.0\columnwidth]{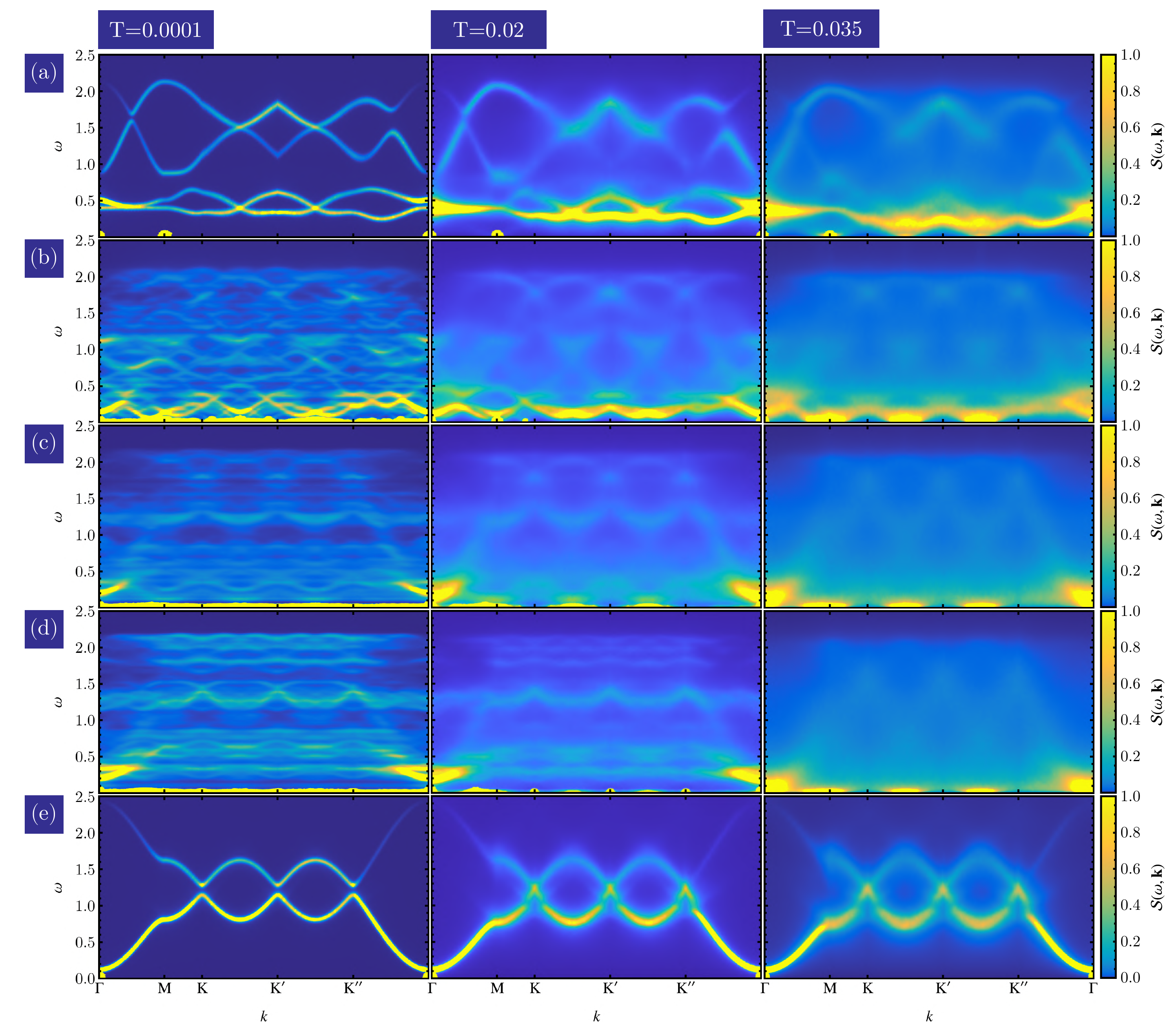}%
	\subfloat{\label{fig:dynamicAll:a}}%
    \subfloat{\label{fig:dynamicAll:b}}%
    \subfloat{\label{fig:dynamicAll:c}}%
    \subfloat{\label{fig:dynamicAll:d}}% 
    \subfloat{\label{fig:dynamicAll:e}}%
	\caption{The dynamical structure factor obtained from LLG simulations is shown for different points of the phase diagram from \cref{fig:phase_diagram_1}, along the $\Gamma = 0.2$ line. The three different columns depict results for three different temperatures (measured in units of the Kitaev exchange $|K|$). For the highest temperature (right column) the zizgag phase (panel (a)) is still ordered but the intermediate field phases' long-range order has disappeared. While the intermediate field phases (panels (b), (c) and (d)) have distinct spin wave excitations at low temperature in their long-range order, at higher temperatures they show a very similar broad scattering continuum up to the magnetic bandwidth $\omega \approx 2.2$. Features at zero frequency may be numerical artifacts, \ie, singularities broadened by a Gaussian frequency filter. The intensities of the different subplots were each individually normalized.}
	\label{fig:dynamicAll}
\end{figure*}
%%%%%%%%%%%%%%%%%%%%%%%%%%%%%%%%%%%%%%%%%%%%%%%%%%%%%%%%%%%%%%
%%%%%%%%%%%%%%%%%%%%%%%%%%%%%%%%%%%%%%%%%%%%%%%%%%%%%%%%%%%%%%

The main objective of our work is to study the dynamical spin structure factors $\mathcal{S} \left( \omega, \bf{k} \right)$ as probed by INS experiments. The latter is defined as the Fourier transform of the dynamic spin-spin correlation function
\begin{equation}
    \label{eqn:dyn_ss_factor}
    \mathcal{S} \left( \omega, \mathbf{k} \right) = \sum_{i,j,\nu}   \int \text{d} t  \  e^{i \mathbf{q} \cdot \left( \mathbf{r}_i -\mathbf{r}_j \right)+ i \omega t} \left\langle S^{\nu}_i (t) S^{\nu}_j (0) \right\rangle .   
\end{equation}
In \cref{eqn:dyn_ss_factor}, $\left\langle \dots \right\rangle$ denotes averaging over different thermodynamic ensembles. In practice however, we compute this average in a single simulation run by using the ergodic theorem
\begin{equation}
    \left\langle S^{\nu}_i (t) S^{\nu}_j (0) \right\rangle  = \frac{1}{T_0} \int_{0}^{T_0} \text{d} t' S^{\nu}_i (t+t') S^{\nu}_j (t')
\end{equation}
for a sufficiently large time window $T_0$. The static spin structure factor $ \mathcal{S} \left( \mathbf{k} \right)$ is related to the dynamic one via
\begin{equation}
    \label{eqn:static_ss_factor}
    \mathcal{S} \left( \mathbf{k} \right) = \frac{1}{2 \pi} \int_{-\infty}^{\infty} \text{d} \omega  \mathcal{S} \left( \omega, \mathbf{k} \right)  .
\end{equation}

We solve the stochastic LLG equation starting from an equilibrated spin configuration and employ a time window of $T_0 = 25 \,000$ divided into $50 \, 000$ time steps (for $T=0.0001$ we extend the time window to $T_0=70 \, 000$ for better frequency resolution).
We have checked that a longer time window and more time steps do not change the results. Consequently, we find that we can calculate the dynamic structure factors without averaging over initial spin configurations as the stochastic field renders the system self-averaging.

A well-known problem of classical spin dynamics calculations is that both Monte Carlo sampling, as well as our stochastic method with a white noise field lead to a classical Boltzmann distribution of excitations. As a result, the weight of the dynamical structure factor over different frequency components differs compared to the correct quantum calculation (in which harmonic spin excitations obey the Bose-Einstein distribution). There are two ways to overcome -- at least partially --  this problem. Within the stochastic LLG approach, one can implement a quantum thermostat via a coloured noise field which fulfills the quantum fluctuation-dissipation theorem leading to the correct Bose-Einstein thermal distribution of the harmonic excitations~\cite{Barker2019}. A numerically much cheaper, albeit more phenomenological, alternative for the dynamic structure factor is to simply rescale the intensity~\cite{zhang2019dynamical,bai2019magnetic}. The key idea is to match the definition of the classical and quantum fluctuation-dissipation relations of the spin structure factor (see Appendix H of Ref.~\cite{smith2022case} for a recent discussion). In this work we use white noise and rescale the numerically calculated dynamic structure factor $\mathcal{S} \left(\omega, \mathbf{k}\right)$ by a factor of $\beta \omega / (1-e^{-\beta \omega})$ reducing the spectral weight at small frequencies for low temperatures. This way of correcting shortcomings of a purely classical calculation has recently been shown to give quantitatively similar results as the $1/S$ Holstein-Primakoff expansion, with qualitative agreement to INS experiments on frustrated three-dimensional magnets ~\cite{zhang2019dynamical,bai2019magnetic,smith2022case}. 

\section{Results}
In \cref{fig:dynamicAll} we show the dynamical spin structure factor for five representative points along the $\Gamma = 0.2$ line of the phase diagram \cref{fig:phase_diagram_1} for three different temperatures (see \cref{app:sec:G05results} for results along the $\Gamma = 0.5$ line).

The low and the high field regimes from \cref{fig:dynamicAll:a,fig:dynamicAll:e} show sharp spin wave excitations in the ordered phases at lowest temperature (left column). In the zigzag phase shown in \cref{fig:dynamicAll:a}, the main intensity is centered around the $\Gamma$ point being contributed by the two lowest-frequency modes. The latter are broadened and only weakly dispersing (as opposed to the two higher frequency branches). As soon as the system enters the field polarized state from \cref{fig:dynamicAll:e}, sharp spin waves appear, which are robust to thermal fluctuations. In contrast, for increasing temperatures the excitations of the zigzag phase broaden significantly. For our choice of highest temperature $T=0.035$ the order parameter of the zigzag phase is significantly reduced by thermal fluctuations, but still nonzero. Nevertheless, spin excitations are very diffusive and the two low frequency modes quickly merge into one broad mode for increasing temperature which is reminiscent of the weakly-dispersive broad mode measured in $\alpha-$RuCl$_3$~\cite{banerjee2016proximate,banerjee2017neutron,do2017majorana}.
We find that above the ordering temperature $T_N \approx 0.04$ the response turns into a scattering continuum over the whole magnetic bandwidth (not shown).

\begin{figure}[t!]
	\centering
	\includegraphics[trim=5 0 25 20, clip, width=1\columnwidth]{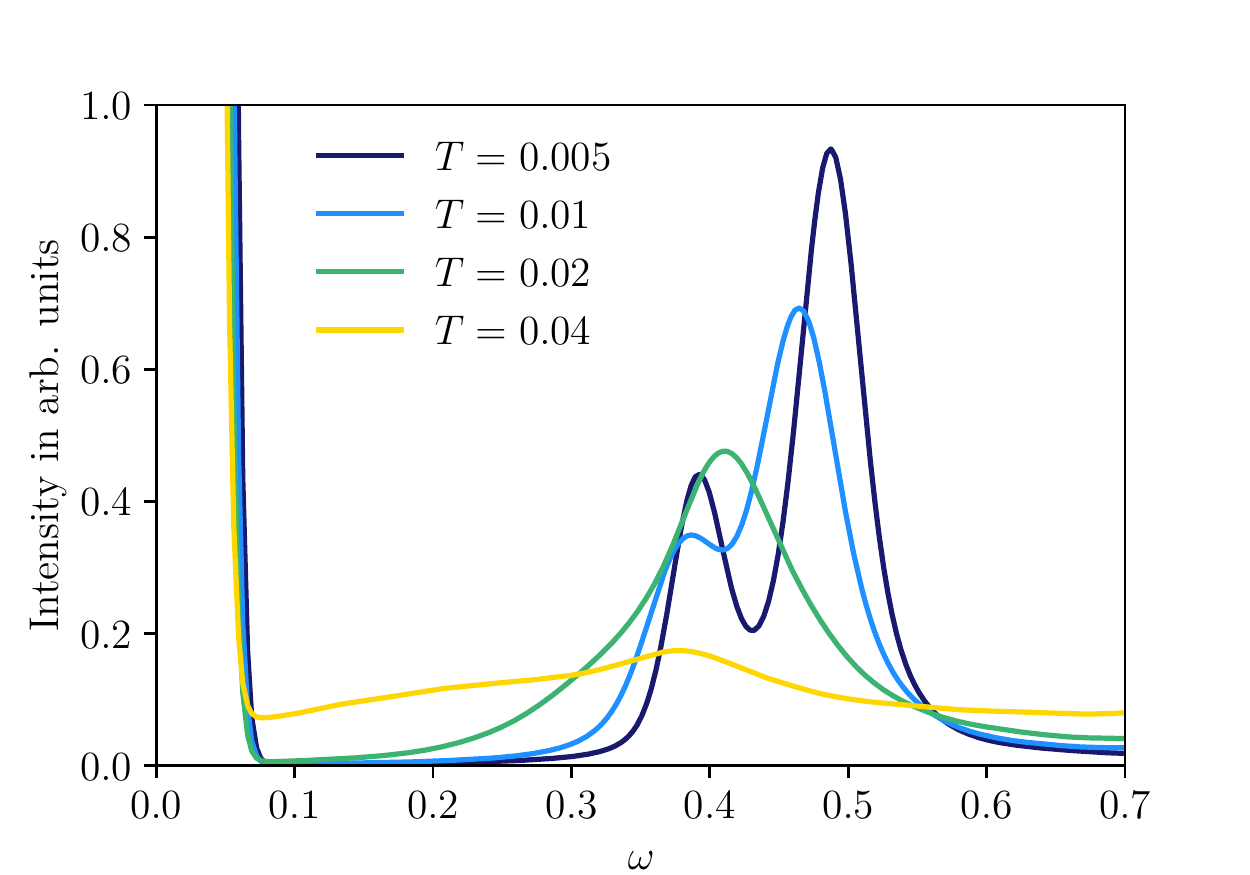}
	\caption{The dynamical response at the $\Gamma$-point of the zigzag phase at four different temperatures. With increasing temperature, the two peaks of the low frequency modes merge into a single mode, which eventually disappears into a broad continuum.}
	\label{fig:dynamicGammaPoint}
\end{figure}

To investigate the thermal broadening of the excitations in the zigzag phase in more detail, we show the response at the $\Gamma$-point for four different temperatures in \cref{fig:dynamicGammaPoint}. Indeed, the two-peak structure of the two spin wave modes quickly merges into one mode even in the ordered regime before disappearing into a broad continuum in the disordered phase.

Next, we turn to the three different intermediate-field-induced phases from \cref{fig:dynamicAll:b,fig:dynamicAll:c,fig:dynamicAll:d}, which show distinct spin-wave excitations in their low-temperature ordered phases. Again, the low-frequency dispersive modes carry most of the intensity. Due to the large unit cells, the intermediate-field phases have a large number of spin-wave branches at high frequency. For increasing temperature, the fragile orders melt well below $T_N$, giving rise to a broad scattering continuum over the entire magnetic bandwidth. Remarkably, the three different fields corresponding to the different intermediate phases display a very similar higher frequency continuum response for temperatures when the order has disappeared. In \cref{fig:dynamicAll:c,fig:dynamicAll:d}, the main intensity is centered around the $\Gamma$-point which, in conjunction with the broad scattering continuum, is again reminiscent of the INS results for $\alpha-$RuCl$_3$~\cite{BAN18}. 

\begin{figure*}[t!]
	\centering
	\subfloat{\label{fig:finiteomega:a}}%
    \subfloat{\label{fig:finiteomega:b}}%
    \subfloat{\label{fig:finiteomega:c}}%
    \subfloat{\label{fig:finiteomega:d}}%
    \subfloat{\label{fig:finiteomega:e}}% 
	\includegraphics[width=2.0\columnwidth]{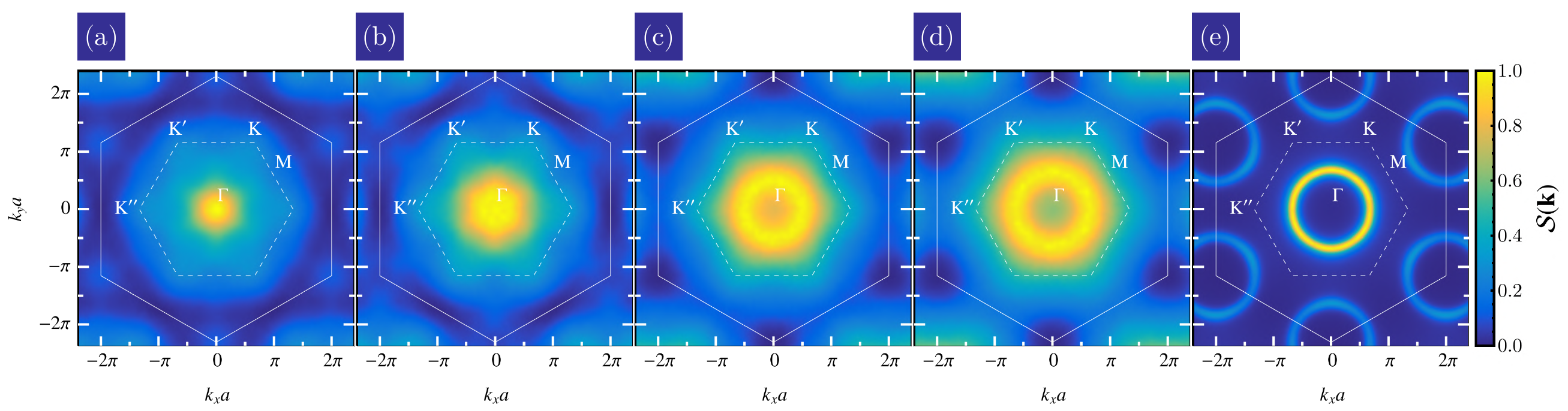}
	\caption{The dynamical response for fixed $\omega=0.4$ is shown at temperature $T=0.035$ for an ensemble average over different configuration orientations of the field induced phases. The broad star-like scattering feature of the zigzag phase (panel (a)), which is reminiscent of the INS results for $\alpha-$RuCl$_3$~\cite{banerjee2017neutron}, turns into a broad ring-like feature for increasing magnetic field. }%
	\label{fig:finiteomega}
\end{figure*}

Finally, we show the dynamical structure for a fixed frequency $\omega=0.4$ at elevated temperature $T=0.035$ in \cref{fig:finiteomega}. In the zigzag phase from \cref{fig:finiteomega:a}, the broad scattering takes the form of a star-like pattern akin to the one found in INS experiments on $\alpha-$RuCl$_3$~\cite{banerjee2017neutron}. For increasing magnetic field, the region of maximum intensity changes from  the $\Gamma$ point to a ring-like shape, which again is very broad in momentum space because of the short real space correlations resulting from the magnetic frustration and thermal disordering. Only in the field polarised state from \cref{fig:finiteomega:e} does the normal sharp ring expected from spin wave excitations reappear. 

%%%%%%%%%%%%%%%%%%%%%%%%%%%%%%%%%%%%%%%%%%%%%%%%%%%%%%%%%%%%%%
%%%%%%%%%%%%%%%%%%%%%%%%%%%%%%%%%%%%%%%%%%%%%%%%%%%%%%%%%%%%%%
%%%%%%%%%%%%%%%%%%%%%%%%%%%%%%%%%%%%%%%%%%%%%%%%%%%%%%%%%%%%%%
%%%%%%%%%%%%%%%%%%%%%%%%%%%%%%%%%%%%%%%%%%%%%%%%%%%%%%%%%%%%%%
%%%%%%%%%%%%%%%%%%%%%%%%%%%%%%%%%%%%%%%%%%%%%%%%%%%%%%%%%%%%%%
%%%%%%%%%%%%%%%%%%%%%%%%%%%%%%%%%%%%%%%%%%%%%%%%%%%%%%%%%%%%%%
%%%%%%%%%%%%%%%%%%%%%%%%%%%%%%%%%%%%%%%%%%%%%%%%%%%%%%%%%%%%%%
%%%%%%%%%%%%%%%%%%%%%%%%%%%%%%%%%%%%%%%%%%%%%%%%%%%%%%%%%%%%%%

\section{Discussion and Conclusion}
We have shown that thermal fluctuations drastically affect the dynamical spin response in frustrated spin models relevant for Kitaev materials like $\alpha-$RuCl$_3$. Already in the absence of any applied magnetic field, the spin-wave excitations of the zigzag state quickly broaden for increasing temperature, even in the ordered low-temperature phase. In contrast, in the high-field spin-polarized phase the spin-wave excitations remain sharp up to high temperature.
Within our semiclassical description, finite temperature fluctuations appear as a stochastic field in the LLG dynamics. Thus, the increased sensitivity of the zigzag phase can be directly traced back to the frustrated interactions of the extended Kitaev model, which allow the stochastic thermal fluctuations to transition between a large number of approximately degenerate spin configurations. 

Most interestingly, we find that the different field-induced intermediate phases from \cref{fig:phase_diagram_1}, are even more fragile with respect to thermal fluctuations: their ordering quickly disappears for experimentally relevant temperatures. The corresponding INS response shows only a broad scattering continuum which only weakly depends on the magnetic field. In connection to $\alpha-$RuCl$_3$, a broad continuum response has been observed at temperatures down to \SI{2}{\kelvin}. For realistic values of \mbox{$K\approx$ \SI{100}{\kelvin}} we find that this is about the same scale as the ordering temperatures of the fragile large-unit-cell intermediate-field phases and thermal broadening effects remain significant. Hence, in order to understand the origin of the INS response, more measurements at lower temperature are highly desirable. 

Our work highlights the importance of thermal fluctuations for accurately describing the INS response of QSL-candidate materials. To distinguish different scenarios of broad scattering -- \ie{} fractionalized excitations of a genuine QSL~\cite{KNO14,punk2014topological}, nonlinearities of magnon-magnon interactions~\cite{winter2017breakdown,zhitomirsky2013colloquium}, or thermal fluctuations between approximately degenerate spin configurations -- requires a careful comparison between different theoretical predictions and experiments at the lowest possible temperatures. Of course, the scenarios are not mutually exclusive, but might be at play simultaneously, which could further complicate the picture.

The general lesson of the present as well as previous~\cite{SAM17,zhang2019dynamical,bai2019magnetic,smith2022case,BHA22} works is that the finite temperature response of a frustrated classical magnet can look surprisingly similar to the one expected from fractionalized excitations in a QSL at zero temperature. This makes the unambiguous observation of quantum fractionalization a challenging task for scattering experiments (at least if they are not performed at temperatures several orders of magnitude below the magnetic exchange scales). On the positive side, the stochastic LLG equation employed here should be a powerful method for comparing different models to INS data at different temperatures in order to extract the microscopic Hamiltonian parameters~\cite{samarakoon2022extraction}. 

In the future, it would be worthwhile to study spin and thermal transport of frustrated (Kitaev) magnets within the stochastic LLG approach, to investigate other dynamical probes like inelastic light scattering, and to explore the effect of quenched disorder. In general, we expect that a clear diagnostic of genuine quantum spin fractionalization will require complementary experimental measurements at lowest temperatures and comprehensive comparison to different quantum as well as semiclassical methods.

\section{Acknowledgements}
We thank P. Roy, R. Otxoa, P. McClarty, R. Moessner, S. Nagler, B. Placke for helpful discussions and especially A. Banerjee also for detailed comments on the manuscript. JK acknowledges support via the Imperial-TUM flagship partnership. The research is part of the Munich Quantum Valley, which is supported by the Bavarian state government with funds from the Hightech Agenda Bayern Plus.
OF acknowledges funding by the Deutsche Forschungsgemeinschaft (DFG, German Research Foundation) – Project-ID 328545488 – TRR 227, project B03.
AN acknowledges funding by the Royal Society.

\bibliographystyle{apsrev4-1}
\bibliography{structFact}

\clearpage
\appendix

\section{Adaptive Runge-Kutta method}
	\label{app:sec:RK_method}
	In this appendix, we provide details on the RK method used to numerically simulate the finite-temperature spin dynamics. The stochastic LLG equation from \cref{eq:LLG} can be rewritten as a generic first-order stochastic differential equation
	\begin{equation}
	    \label{app:eqn:generic_stoch_eqn}
	    \mathrm{d} \mathbb{S}=f\left(\mathbb{S},t\right)\mathrm{d} t+g\left(\mathbb{S},t\right) \mathrm{d}\mathbb{W},
	\end{equation} 
	where $\mathbb{S}$ denotes the vector containing all the components of the spins in the lattice, and $\mathbb{W}$ represents a corresponding vector of Wiener processes. The explicit forms of vector functions $f\left(\mathbb{S},t\right)$ and $g\left(\mathbb{S},t\right)$ can be determined directly from \cref{eq:LLG}. For the purpose of explaining the RK numerical integration method, we leave their forms unspecified. The spin values at a given time-step $t_{n+1}$ can be obtained from the ones from the previous time-step $t_{n}$ by employing a RK approximation of order $p=4$
	\begin{equation}
		\mathbb{S}_{n+1} = \mathbb{S}_{n} + \sum^{p}_{i=1} b_i \mathbb{K}_i + \dfrac{1}{2} \left( \eta_1 + \eta_2 \right), 
	\end{equation}
	where we have defined
	\begin{align}
			\eta_1&= g\left(\mathbb{S},t_n\right) \sqrt{\Delta t_n},\nonumber\\
			\mathbb{K}_1 &=  f\left(\mathbb{S},t_n\right) \Delta t_n,\nonumber\\
			\eta_2&= g\left(\mathbb{S}+\mathbb{K}_1 + \eta_1,t_n+\Delta t_n\right) \sqrt{\Delta t_n},\nonumber\\
			\mathbb{K}_i &=  f\left(\mathbb{S} + c_i \eta_1 +  \sum^{i-1}_{j=1}a_{ij} \mathbb{K}_j, t_n + c_i \Delta t_n \right) \Delta t_n, i>1, \label{AP4}
	\end{align}
	and $\Delta t_n=t_{n+1}-t_{n}$. The coefficients $a_{ij}$, $b_i$ and $c_i$ are tabulated and depend on the specific RK method used. Eq. \eqref{AP4} represents a generalization of the RK4-Heun~\cite{AME16} method, where the deterministic part of the differential equation is computed by a $p$-th order RK method. Because the stochastic term of the equation is evaluated at both ends of the interval, the result will converge to the Stratonovich solution~\cite{AME16}.
	
	In order to obtain an estimate for the error after each time-step, the $p$-th order RK method is complemented by another one of order $p+1$. The coefficients $a_{ij}$, $b_i$ and $c_i$ corresponding to each method are chosen in such a way as to ensure a minimal number of evaluations of the function $f$. This is achieved by using the same set of intermediate values $\lbrace \mathbb{K}_i \rbrace$. Once an estimate of the error ($\mathbb{E}_{n+1}$) is found, the next time interval, $\Delta t_{n+1}$, can be adjusted according to~\cite{DOR80}
	\begin{equation}
		\Delta t_{n+1}=0.9\left(\dfrac{\delta}{\mathbb{E}_{n+1}}\right)^{\frac{1}{p}+1}.
	\end{equation} 
	This ensures that the algorithm uses the largest time interval that keeps the truncation error below a given tolerance, $\delta$. For the specific case of the sLLG equation, we employ the method RK5(4)7S~\cite{DOR80}, that was adapted according to Eq. \eqref{AP4} to account for the stochastic magnetic field.

\section{Simulated annealing procedure}
	\label{app:sec:sim_annealing}	
	In this appendix, we outline the procedure used to determine the classical ground states of the system used as initial conditions for computing the static and dynamic structure factors. It should be noted that the simulated annealing procedure is necessary precisely because using a random spin configuration as an initial condition at low temperatures will trap the system into a meta-stable state.
	
	To prevent the formation of such meta-stable states, we start from a temperature much higher than the ordering temperatures and then progressively cool the system until the target temperature (\ie{} the temperature at which we simulate the spin structure factors) is reached. There are multiple options one can employ for the cooling protocols. In this work, we repeatedly linearly cool and slightly reheat the system until we reach the target temperature. At each cooling step we encourage the system to settle in the lowest-energy state. The reheating steps destroy any meta-stable states that could form during the cooling stage.
	
	Our annealing procedure starts from an initial temperature  $T_{\mathrm{init}}$ and consists of $N_{\mathrm{a}}$ cycles each lasting a time $\Delta t_{\mathrm{a}}$. During one cycle, the temperature is first linearly decreased by a factor $f_{\mathrm{c}}$ during a time $\Delta t_{\mathrm{h}}$ (we use the convention in which $f_{\mathrm{c}} < 1$ denotes a net cooling). The cooling stage is immediately followed by a reheating step in which the temperature is raised by a factor $f_{\mathrm{h}}>1$ (with $f_{\mathrm{h}} f_{\mathrm{c}}<1$) in a time $\Delta t_{\mathrm{h}}$ (such that $\Delta t_{\mathrm{a}} = \Delta t_{\mathrm{c}} + \Delta t_{\mathrm{h}}$). As such, the temperature evolution is given by
	\begin{equation}
		\frac{T \left( n \Delta t_{\mathrm{a}} + t \right)}{T_{\mathrm{init}}} = \left( f_{\mathrm{h}}f_{\mathrm{c}} \right)^n \left[ \frac{t \left( f_{\mathrm{c}} - 1 \right) }{\Delta t_{\mathrm{c}}} + 1 \right],
	\end{equation}
    for $0\leq t \leq \Delta t_{\mathrm{c}}$, and $0 \leq n < N_a $ ($n \in \mathbb{Z}$), during the cooling stage. Similarly, during the heating stage 
	\begin{equation}
		\frac{T \left( n \Delta t_{\mathrm{a}} + \Delta t_{\mathrm{c}} + t \right)}{T_{\mathrm{init}}} = \left( f_{\mathrm{h}}f_{\mathrm{c}} \right)^n f_{\mathrm{c}} \left[ \frac{t \left( f_{\mathrm{h}} - 1 \right) }{\Delta t_{\mathrm{h}}} + 1 \right],
	\end{equation}
	for $0\leq t \leq \Delta t_{\mathrm{h}}$. For the simulations, we use \mbox{$N_{\mathrm{a}} = 45 $} cycles with $f_{\mathrm{c}} \approx 0.1$, \mbox{$f_{\mathrm{h}} \approx 8$}, \mbox{$\Delta t_{\mathrm{c}} \approx 200 \, 000$}, and \mbox{$\Delta t_{\mathrm{h}} \approx 20 \, 000$}. The temperature effectively decreases by a factor \mbox{$f_{\mathrm{h}} f_{\mathrm{c}}<1$} during each cycle.
	
\section{Results for $\Gamma = 0.5$}
\label{app:sec:G05results}
%%%%%%%%%%%%%%%%%%%%%%%%%%%%%%%%%%%%%%%%%%%%%%%%%%%%%%%%%%%%%%
%%%%%%%%%%%%%%%%%%%%%%%%%%%%%%%%%%%%%%%%%%%%%%%%%%%%%%%%%%%%%%
\begin{figure*}[t]
	\centering
	\includegraphics[width=2.0\columnwidth]{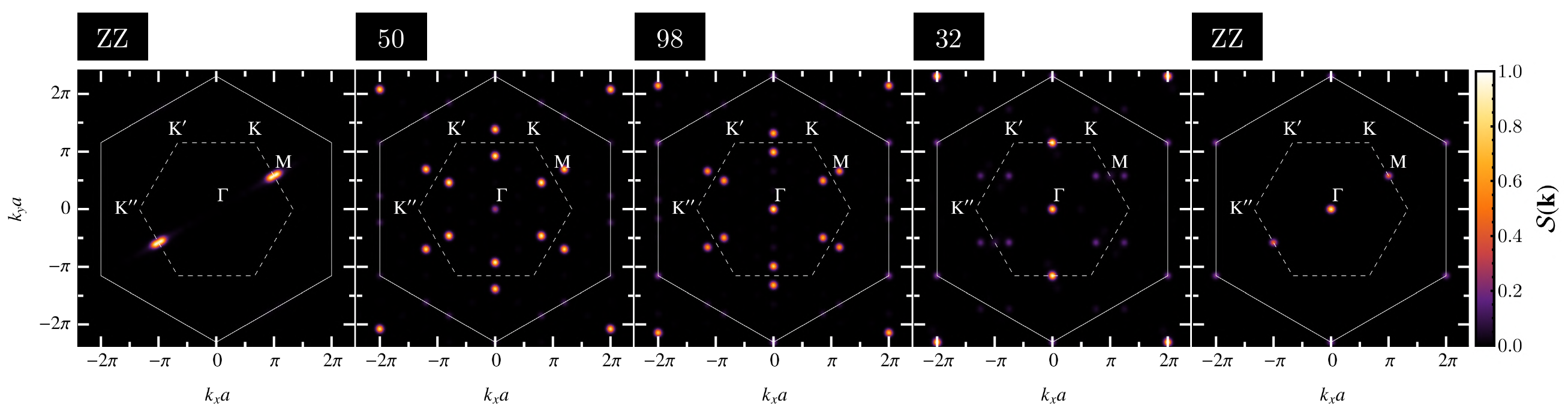}
	\caption{Static structure factors along the $\Gamma = 0.5$ line of the phase diagram in \cref{fig:phase_diagram_1} at low temperature $T=0.0001$. The sharp peaks are broadened by a Gaussian filter for visibility and the first Brillouin zone is indicated by a dashed line. We show results for the points $|\mathbf{h}| \in \{0.01,0.3,0.682,0.86,1.108\}$ with $\Gamma^\prime = -0.02$. For higher magnetic fields $|\mathbf{h}|>1.29$ we obtain the polarized ferromagnetic state.}
	\label{fig:staticAllG05}
\end{figure*}
\begin{figure}[h]
	\centering
	\includegraphics[width=0.9\columnwidth]{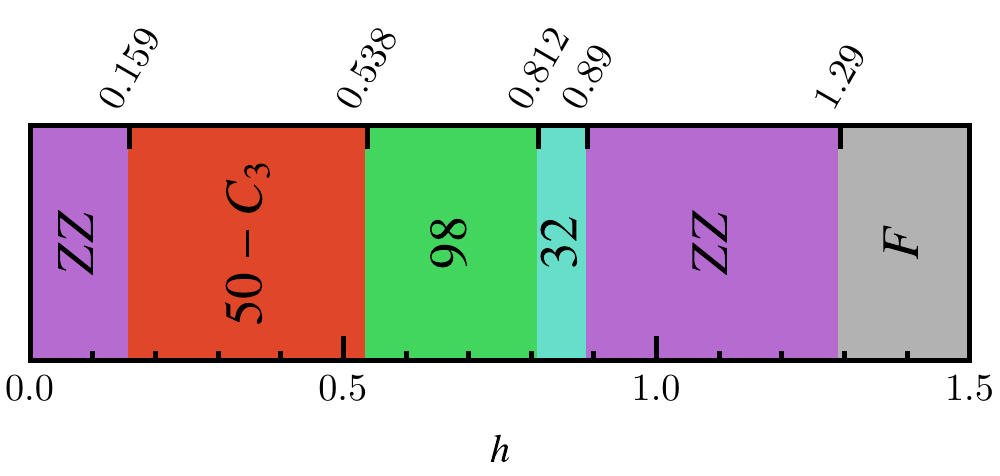}%
	\caption{Detailed phase diagram of the $\Gamma=0.5$ line in \cref{fig:phase_diagram_1}, as obtained by Ref.~\cite{CHE20a}. The various phases are labeled as in \cref{fig:phase_diagram_1}.}
	\label{fig:phase_diagram_2}
\end{figure}

%%%%%%%%%%%%%%%%%%%%%%%%%%%%%%%%%%%%%%%%%%%%%%%%%%%%%%%%%%%%%%
%%%%%%%%%%%%%%%%%%%%%%%%%%%%%%%%%%%%%%%%%%%%%%%%%%%%%%%%%%%%%%
Analogous to the results of the main text, in this appendix, we show the static and dynamic structure factors at different temperatures along the $\Gamma = 0.5$ line of the phase diagram from \cref{fig:phase_diagram_1}, shown in more detail in \cref{fig:phase_diagram_2}. We obtain all structure factors except the first zigzag phase ($h = 0.01$), as described in \cref{app:sec:sim_annealing}. For the latter, we employ the high-field zigzag phase ($h = 1.108$) as the initial configuration and do not use an annealing procedure. For the low-field zigzag phase, we find that the spin configuration obtained through simulated annealing is modulated and features additional low-intensity spin waves. As for the $\Gamma = 0.2$ line, we identify the different orders via their real space spin configurations and by their static structure factors, which are shown in \cref{fig:staticAllG05}.

%%%%%%%%%%%%%%%%%%%%%%%%%%%%%%%%%%%%%%%%%%%%%%%%%%%%%%%%%%%%%%
%%%%%%%%%%%%%%%%%%%%%%%%%%%%%%%%%%%%%%%%%%%%%%%%%%%%%%%%%%%%%%
%%%%%%%%%%%%%%%%%%%%%%%%%%%%%%%%%%%%%%%%%%%%%%%%%%%%%%%%%%%%%%
%%%%%%%%%%%%%%%%%%%%%%%%%%%%%%%%%%%%%%%%%%%%%%%%%%%%%%%%%%%%%%

\begin{figure*}[t!]
	\centering
	\includegraphics[width=1.9\columnwidth]{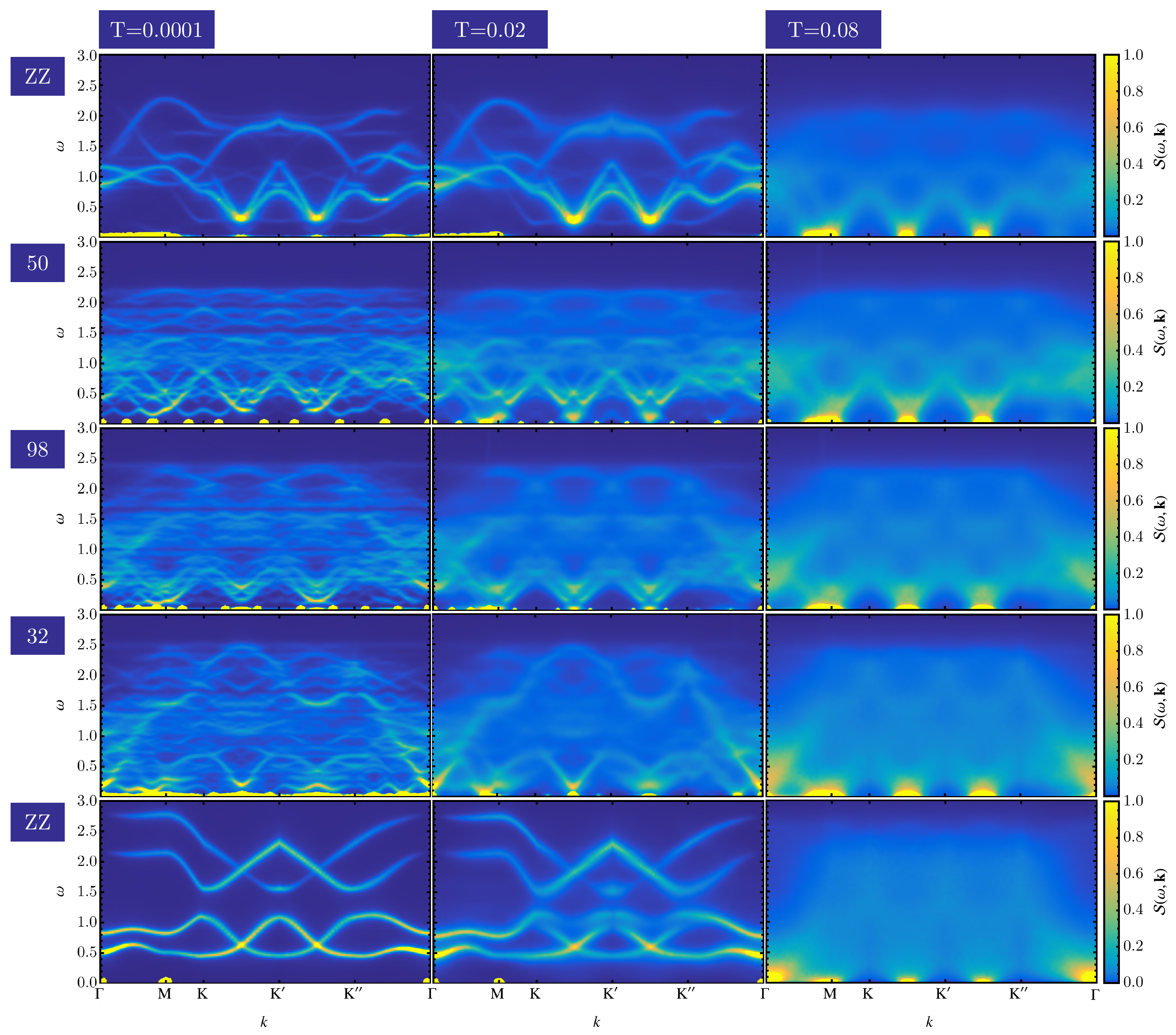}%
	\caption{The dynamical structure factor obtained from LLG simulations is shown for different points of the phase diagram, see \cref{fig:phase_diagram_2} and \cref{fig:phase_diagram_1} along the $\Gamma = 0.5$ line. The three different columns depict results for three different temperatures (measured in units of the Kitaev exchange $|K|$). The intermediate orders host a multitude of distinct spin wave excitations and are distinguishable up to higher temperatures than the orders along the $\Gamma = 0.2$ line. Since the low-field ZZ phase is less stabilized by an external field, some additional low-intensity modes from domain walls appear in the simulation that are not inherent features of the ZZ order.}
	\label{fig:dynamicAllG05}
\end{figure*}

We observe a thermal broadening of the spin wave excitations, which are, however, more stable against thermal fluctuations than before, due to the relatively stronger off-diagonal exchange interaction. Whereas the intermediate phases in \cref{fig:dynamicAll} at temperature $T=0.035$ show a similarly broad (almost) continuum response, the different phases for $\Gamma=0.5$ are clearly distinguishable at least up to $T=0.05$ (not shown here). For even higher temperatures, however, the dynamic response of all phases again looks very similar as is shown in the right panel of \cref{fig:dynamicAllG05}. We further note that the magnetic bandwidth increases with stronger magnetic field from $h \approx 2.3$ to $h \approx 2.5$ for the 32-site order.

Thermal fluctuations have a particularly interesting effect on both the 50-site and 98-site order. With increasing temperature, the modes around the K-points become symmetric, which might be attributed to different magnetic domain realizations, and the temperature causes a split of the low frequency mode at the M-points on a path between K-points. The similarity between these two orders is not surprising because the 98-site order appears like an augmented 50-site order, as already observed in Ref. \cite{CHE20a}.

%%%%%%%%%%%%%%%%%%%%%%%%%%%%%%%%%%%%%%%%%%%%%%%%%%%%%%%%%%%%%%
%%%%%%%%%%%%%%%%%%%%%%%%%%%%%%%%%%%%%%%%%%%%%%%%%%%%%%%%%%%%%%
%%%%%%%%%%%%%%%%%%%%%%%%%%%%%%%%%%%%%%%%%%%%%%%%%%%%%%%%%%%%%%
%%%%%%%%%%%%%%%%%%%%%%%%%%%%%%%%%%%%%%%%%%%%%%%%%%%%%%%%%%%%%%

\begin{figure*}[h]
	\centering
	\includegraphics[width=2.0\columnwidth]{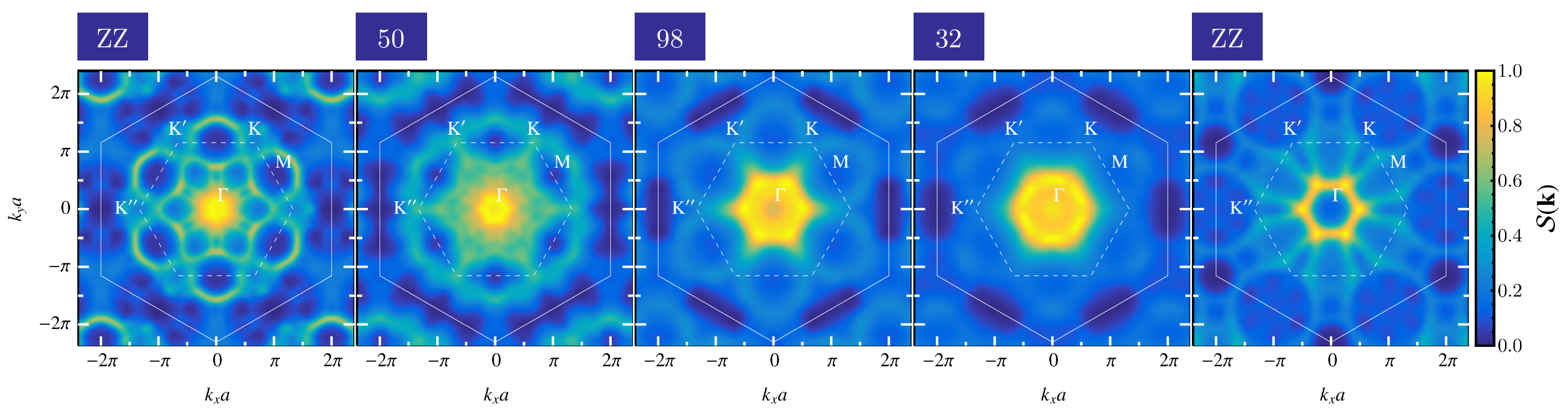}%
	\caption{The dynamical response for fixed $\omega=0.9$ is shown at temperature $T=0.035$ for an ensemble average over different configuration orientations of the field induced phases. As in \cref{fig:finiteomega}, we reproduce a star-like scattering feature for the ZZ phases.}
	\label{fig:finiteomegaG05}
\end{figure*}

The dynamic response at fixed \mbox{$\omega = 0.9$} seen in \cref{fig:finiteomegaG05} again resembles the star-shaped scattering feature expected from INS experiments of $\alpha-$RuCl$_3$ \cite{banerjee2017neutron} and we observe the ring-like feature around the $\Gamma$-point for increasing fields.

%%%%%%%%%%%%%%%%%%%%%%%%%%%%%%%%%%%%%%%%%%%%%%%%%%%%%%%%%%%%%%
%%%%%%%%%%%%%%%%%%%%%%%%%%%%%%%%%%%%%%%%%%%%%%%%%%%%%%%%%%%%%%
%%%%%%%%%%%%%%%%%%%%%%%%%%%%%%%%%%%%%%%%%%%%%%%%%%%%%%%%%%%%%%
%%%%%%%%%%%%%%%%%%%%%%%%%%%%%%%%%%%%%%%%%%%%%%%%%%%%%%%%%%%%%%

\end{document}